\documentclass[prb,twocolumn,showpacs,amssymb]{revtex4}

\usepackage{graphicx}
\usepackage{ifthen}
\usepackage{ifpdf}

\usepackage{latexsym}
\usepackage{amssymb}
\usepackage{bm}
\newcommand{\half}{\mbox{\small $\frac{1}{2}$}}
\newcommand{\pd}[2]{\frac{\partial #1}{\partial #2}}

\newcommand{\eexp}{\mbox{e}^}

\newcommand{\rr}{{\bf r}}
\newcommand{\qq}{{\bf q}}
\newcommand{\uu}{{\bf u}}
\newcommand{\vv}{{\bf v}}
\newcommand{\EE}{{\bf E}}
\newcommand{\text}[1]{\mbox{#1}}

\newcommand{\kpar}{k}

\newcommand{\beq}[1]{\begin{eqnarray}\ifthenelse{#1=-1}{\nonumber}
{\ifthenelse{#1=0}{}{\label{e#1}}}}
\newcommand{\eeq}{\end{eqnarray}}
\newcommand{\hide}[1]{}

\begin{document}

\author{Baruch Horovitz{$^1$} and Carsten Henkel{$^2$} }
\affiliation{{$^1$} Department of Physics, Ben
Gurion University, Beer Sheva 84105, Israel}
\affiliation{{$^2$}Institut f\"{u}r Physik und Astronomie,
Universit\"{a}t Potsdam, 14476 Potsdam,
Germany}


\title{Surface plasmons at composite surfaces with diffusive charges}

\begin{abstract}
Metal surfaces with disorder or with nanostructure modifications are studied, allowing for a localized charge layer (CL) in addition to continuous charges (CC) in the bulk, both charges having a compressional or diffusive non-local response. The
notorious problem of ``additional boundary conditions'' 
is resolved with the help of a Boltzmann equation that involves the scattering between the two charge types. Depending on the strength of this scattering, the oscillating charges can be dominantly CC or CL; the surface plasmon (SP) resonance acquires then a relatively small linewidth, in agreement with a large set of data. With a few parameters our model describes a large variety of SP dispersions corresponding to observed data.
\end{abstract}



\maketitle

\paragraph*{{\bf Introduction.}}

Collective electronic excitations in metal surfaces covered with adsorbates or nanostructures are of significant recent interest. Surface plasmons (SP) are an efficient tool for characterizing such surfaces \cite{garcia,pitarke} and can be used as sensitive chemical sensors and biosensors \cite{homola}.

A considerable amount of data on the dispersion of SPs has been accumulated \cite{sutto,rocca1,tsuei,sprunger,rocca2,savio1,chiarello,savio2,yu,politano1,politano2,politano3,politano4} on a variety of metal surfaces, clean, sputtered, or covered with thin films. These composite surfaces indicate the necessity of distinguishing between two types of charge carriers: continuous charges (CC) extending throughout the bulk and a charge layer (CL) of carriers localized at the surface. In fact, photo\-emission data on some of these surfaces reveals the existence of quantum well states at the surface \cite{varykhalov}. The two charge types are relevant also for pure metals: in alkali metals the charge extends beyond the neutralizing ions, forming a distinct layer \cite{tsuei,sprunger}. In Ag a two-component s-d electron system with different surface and bulk charges has
been put forward to explain the SP dispersion \cite{liebsch}. In some cases a distinct surface band is formed \cite{pitarke}, e.g. as in Be(0001), showing an acoustic plasmon \cite{diaconescu}.
Further motivation for a two-type charge model comes from studies of
the anomalous heating of cold ions observed in miniaturized Paul traps,
that invoke surface charge fluctuations on the metallic
electrodes \cite{Turchette00a, Leibrandt07b, hh, Dubessy09, Daniilidis11}.

Most information about the SP dispersion $\omega( \kpar )$ is available in
the non-retarded range $\omega_p / c \ll \kpar \ll k_F$ where
$\kpar$ is the momentum parallel to the surface,
$\omega_p$ is the
bulk plasma frequency, $c$ the speed of light, and $k_F$ the Fermi momentum.
In this range, the dispersion is parameterized as
$\omega( \kpar ) = A + B \kpar + C \kpar^2$ and
the limiting value $A = \omega_p/\sqrt{2}$ is well known
(assuming unit background permittivity) \cite{ritchie}.
Considerable insight is gained by Feibelman's sum rule \cite{feibelman} relating
the slope $B$ to the centroid of the oscillating charge density profile $\delta n(z)$:
\begin{equation}
	B = - \frac{ \omega_p }{ 2 \sqrt{ 2 } }
	\frac{ \int \! {\rm d}z \, z\,\delta n(z)}{\int\!{\rm d}z \,\delta n(z) }
	.
	\label{eq:Feibelman-slope}
\end{equation}
Here the medium is located in the half-space $z \le 0$, and the onset
of the dielectric function due to bulk ions is 
 at $z = 0$.
Hence the SP dispersion is negative ($B < 0$) if the oscillating charge is
dominantly located outside
the metal ($z > 0$) as in alkali metals \cite{tsuei,sprunger}, while $B > 0$ if the oscillating charge is dominantly inside. In some cases the coefficient $B$ is small and the quadratic term $C \kpar^2$ dominates the dispersion \cite{sutto,politano2,politano4}, even at the lowest measurable $\kpar$. From Eq.(\ref{eq:Feibelman-slope}), $B=0$
is a strong indication for a charge excitation that is highly localized
at the surface, $z=0$.

In the present work we consider a non-local electrodynamic model including both CC and CL, which for small $\kpar$ involves diffusive (or compressional) terms, similar to hydrodynamic models \cite{garcia,pitarke}. The presence of two types of charges leads to the situation that the boundary conditions of Maxwell's equations are not sufficient to solve the problem. The problem has been originally identified by Pekar \cite{pekar}, with a variety of ``additional boundary conditions" (ABC) proposed over the years \cite{zeyher,schwartz,schneider,silveirinha}, and compared with experimental data \cite{schneider}. In Ref.~\onlinecite{forstmann},
a model with two charge types, similar to ours, was considered and an
ABC was proposed by arguing that dissipated energy is conserved across a boundary. It is known, however, from ABC studies \cite{pekar,zeyher,schwartz,schneider,silveirinha} that microscopic information must be used, i.e. Maxwell's equations by themselves are insufficient.

We address here the ABC problem by solving a Boltzmann equation, allowing for impurity scattering in the bulk as well as surface scattering that mixes CC and CL. Note also that the CL allows charge conservation
to hold even if the bulk current perpendicular to the surface is locally finite,
representing deviations from specular reflection. In this sense the localized charge is a measure of diffuse scattering or disorder at the surface: the scattered bulk current becomes a surface current which is allowed to diffuse along the surface. A key point of our model are the transition rates between the two types of charges that provide
a restoring force (plasma oscillation)
and damping for the combined charge oscillations.
By allowing for bulk and surface impurity scattering, our approach complements traditional band structure theories \cite{garcia,pitarke} and provides a unified framework for analyzing SP dispersion and linewidth. The resulting requations involve a few parameters of the poorly known CL and of the surface-bulk charge scattering. We start with discussing the qualitative effect of the surface scattering and then derive the actual SP dispersion, showing a variety of forms corresponding to experimental data.

\begin{table*}
\begin{center}
\begin{tabular}{|c|c|c|c|c|c|}
\hline
Ref. & sample & A (eV) & B (eV$\cdot{\rm \AA}$) & C (eV$\cdot{\rm \AA}^2$) & $\Gamma$ (eV)  \\ \hline\hline
\onlinecite{tsuei} & K & 2.73 & $-0.95$ & 0.09 & 0.3\\ \hline
\onlinecite{sutto} & Ag(110) ${\bf k}\parallel$[001] & 3.76 & -- & 3.68 & 0.11 \\ \hline
\onlinecite{sutto} & Ag(110) ${\bf k}\parallel [1{\bar 1}0]$ & 3.86 & -- & 1.46 & 0.18 \\ \hline
\onlinecite{sutto} & Ag(111)  & 3.69 & -- & 4.17  & 0.06 \\ \hline
\onlinecite{savio1} & 0.05ML O/Ag(001) & 3.71 & -- & 3.1 & $\sim 0.3$ \\ \hline
\onlinecite{savio1} & 0.1ML O/Ag(001) & 3.70 & 1.1 & $-0.9$ & $\sim 0.3$ \\ \hline
\onlinecite{savio1} & 0.15ML O$_2$/Ag(001) & 3.69 & 0.86 & $-0.1$ & $\sim 0.3$ \\ \hline
\onlinecite{savio2} & Ag(001) & 3.71 & 1.45 & -- & 0.15 \\ \hline
\onlinecite{politano1} & Cu(111) & 1.18 & $-0.72$ & -- & 0.2 \\ \hline
\onlinecite{politano2} & 10ML Ag/Ni(111) & 3.75 & -- & 1.57 & $\sim 0.1$ \\ \hline
\onlinecite{politano4} & sputtered 22ML Ag/Cu(111) & 3.76 & $-0.08$ & 2.50 & 0.18 \\ \hline
\onlinecite{sprunger} & Mg(0001) & 7.38 & $-3.023$ & 9.78 & 1.2 \\ \hline
\onlinecite{chiarello} & Al(111) & 10.86 & $-3.1$ & 7.7 & 2.3 \\ \hline
\end{tabular}
\end{center}
\caption{Data on SP dispersion $\omega=A+B \kpar+C \kpar^2$ and linewidth $\Gamma$ at small $\kpar$. The coefficients $A,B,C$ are given by the corresponding references, with ``--'' indicating a coefficient that was assumed to be zero for the fit. All cases with
$|B| > C k_1$ or $|B| < 0.1\,C k_1$ are shown with a relatively small $\Gamma$
(momentum scale $k_1 = 1\,{\rm \AA}^{-1}$).
The last two entries show cases where  $0.1\,C k_1 < |B| < C k_1$ with a large
$\Gamma$.}
\label{t:table-with-SPdata}
\end{table*}

\paragraph*{\bf{Qualitative features.}}
Our model implies that the surface scattering that mixes CC and CL enhances
the SP linewidth $\Gamma$. Hence $\Gamma$ is relatively small if the charge
response is dominantly CC
(large positive slope, $B > C k_1$
with $k_{1} = 1\,{\rm \AA}^{-1}$ being a typical large momentum)
or dominantly CL
(large negative slope, $B < - C k_1$,
or small slope, $|B| < 0.1 \, C k_1$).
Table~\ref{t:table-with-SPdata} shows all the corresponding cases that we are aware of, confirming this trend for small damping.
Cases where $0.1\, C k_1 < |B| < C k_1$
may or may not correspond to dominant CC or CL and a detailed fit of
$\omega( \kpar )$ and $\Gamma( \kpar )$ is needed to determine the parameters.


\paragraph*{{\bf The model and nonlocal electrodynamics.}}

We proceed to define the charges and currents of the model. The bulk
medium is located in $z < 0$, and the surface layer in $0 < z < d$.
We write $\rho( {\bf r} )$ and $\rho_s( {\bf r} )$, respectively,
for the charge density in the two regions.
We expand the current response, considering small charge gradients
\beq{01}
\frac{1}{\tau}{\bf j} + \partial_t{\bf j} &=&
\frac{\omega_p^2}{4\pi}{\bf E} - c_b^2{\bm\nabla}\rho
	\qquad z<0
	\nonumber
\\
\frac{1}{\tau_s}{\bf j}_s + \partial_t{\bf j}_s &=&
\frac{\omega_s^2}{4\pi}{\bf E} - c_s^2{\bm\nabla}\rho_s
	\qquad 0<z<d
	\,.
\eeq
The equilibrium charge density provides restoring forces $\omega_p^2/4\pi$
and $\omega_s^2/4\pi$ in the bulk and surface regions, where $\omega_{p,s}$
are the corresponding plasma frequencies if the CC/CL were decoupled. The bulk and surface compressibilities
are expressed in terms of the sound velocities $c_{b,s}$, respectively,
while $c_b^2 \tau$ and $c_s^2 \tau_s$ are the corresponding diffusion
coefficients.
An important realization of a surface layer is the case of a charge
spill-out, as in the alkali metals \cite{tsuei, sprunger,chiarello}. In this case,
the CL is further away from the restoring force provided by the
equilibrium situation. Hence the equilibrium charge, when averaged over
the CL thickness $d$ in the oscillating state, is reduced and the
restoring force is weak, $\omega_s < \omega_p$. Note that even
with $\omega_s = 0$, there is eventually a restoring force via coupling
to the bulk so that $\omega( \kpar \to 0 ) = \omega_p / \sqrt{ 2 }$.
We also note that when $\omega_s = 0$, Feibelman's sum rule
Eq.(\ref{eq:Feibelman-slope}) \cite{feibelman}
%
is still valid with $z=0$ as origin, since the charges at $z>0$ respond only
in a nonlocal way via the gradient
${\bm\nabla}\rho_s$ which affects higher order $O(\kpar^2)$ terms.
When $\omega_s \ne 0$, e.g. for nanostructures with their own ions
and equilibrium charge density, the sum rule~(\ref{eq:Feibelman-slope})
is modified such that
\begin{equation}
	B  =  \frac{ \omega_pÊ}{ 2\sqrt{2} } \left[
	-
	\frac{ \int \! {\rm d}z \, z\,\delta n(z)}{\int\!{\rm d}z \,\delta n(z) }
	+
	\frac{ \omega_s^2 d }{ \omega_p^2 }
	\right]
\,,
	\label{eq:shifted-slope}
\end{equation}
adding a positive term to the slope.


In the following we omit the damping terms ($1/\tau,1/\tau_s$) for brevity.
They are easily restored by multiplying the bulk [surface] parameters
$\omega_p^2$ and $c_b^2$ [$\omega_s^2$ and $c_s^2$]
with the factor
$1+ i /(\omega\tau)$ [$1 + i /(\omega\tau_s)$], respectively.

We look for solutions to the fields at frequency $\omega$ that vary with
the wave vector $\kpar$ parallel to the surface.
Charge conservation
then determines the charge profiles
\begin{eqnarray}
\label{e02}
z < 0: &&
[\omega_p^2-c_b^2\nabla^2 -\omega^2]\rho(\rr) = 0
	\nonumber\\
&\Rightarrow&
\rho(\kpar,z) = \rho_0(\kpar)\,\eexp{v_bz},
\\
0 < z < d: &&
{}[\omega_s^2-c_s^2\nabla^2-\omega^2]\rho_s(\rr) = 0
	\nonumber\\
&\Rightarrow&
\rho_s(\kpar,z)
= \gamma_0(\kpar)\cosh v_sz
+ \gamma_1(\kpar)\sinh v_sz
	\,.
\nonumber
\end{eqnarray}
The inverse decay lengths
$v_b$ and $v_s$ for the bulk and surface charges are
\beq{03}
v_s&=&\frac{1}{c_s}\sqrt{\omega_s^2+c_s^2 \kpar^2-\omega^2}
	\,,
\nonumber\\
v_b&=&\frac{1}{c_b}\sqrt{\omega_p^2+c_b^2 \kpar^2-\omega^2}
	\,.
\eeq
Current conservation at $z = d$ yields
$(\omega_s^2/4\pi) E_z( \kpar, d) - c_s^2 \partial_z\rho_s( \kpar, d)=0$,
while at $z = 0$,
\beq{04}
-i\omega j_z^{in}&\equiv&\frac{\omega_p^2}{4\pi}E_z( \kpar, 0^-)-c_b^2\partial_z\rho( \kpar, 0^-)
	\nonumber\\
&=&\frac{\omega_s^2}{4\pi}E_z( \kpar, 0^+)-c_s^2\partial_z\rho_s( \kpar, 0^+)
\eeq
where $0^{\pm}$ denotes the limit $z\rightarrow\pm 0$.

Maxwell's equation
determines the longitudinal components of $\bm{E}$ in terms of the charges, however the transverse components require four unknowns: two
amplitudes for upward and downward propagation in the layer, and one amplitude
each in bulk ($z < 0$) and in vacuum ($z > d$).
Together with the amplitudes $\rho_0, \gamma_0, \gamma_1$ for the
charge density, we have seven unknowns. Each interface 
provides three relations (the matching of $E_z,\,\partial_z E_z$ and $j_z$),
hence one boundary condition is missing. The necessity for an additional
boundary condition (ABC) has a long history and occurs when more than
one material mode is present, for example
the $\gamma_0,\gamma_1$ modes of Eq. (\ref{e02})
 \cite{zeyher,schwartz,schneider,silveirinha}.
Therefore a microscopic input is needed to provide the ABC.

Before developing the ABC, we discuss some general properties of the
CC+CL system.
At very small parallel momentum,
$k\lesssim\omega_p/c \sim 2\cdot 10^{-3}\,{\rm \AA}^{-1}$,
the SP dispersion approaches the light dispersion $\omega( \kpar ) \to c \kpar$.
While this range can be optically probed \cite{lirtsman}, most of the data is found by electron energy loss spectroscopy at higher $\kpar$ values. Therefore, we focus on the non-retarded region $\omega_p / c \ll \kpar$,
formally taking $c\rightarrow\infty$.
Since the limit
$\omega( \kpar\rightarrow 0 ) = \omega_p / \sqrt{2}$ is a bulk
property,
we expect that
the parameter $\omega_s$ has a small effect.
In fact, the numerical solutions shown below confirm this for
$\omega_s\lesssim 0.3\,\omega_p$ and $k \lesssim 0.4\,\omega_p/c_b$.
Hence we display the result for
$\omega_s=0$, which has a relatively simple form:
\begin{eqnarray}
\label{e05}
&&
\frac{c_b^2}{\omega_p^2}\rho_0( \kpar )[v_b+\frac{\omega^2v_b-\omega_p^2 \kpar}{\omega^2-\omega_p^2}]=
\\
&&
\frac{c_s^2}{\omega^2}\gamma_0( \kpar )[\kpar(\frac{\eexp{-\kpar d}}{\cosh v_s d}-1)
- v_s(2\frac{\omega^2}{\omega_p^2}-1) \tanh (v_s d)]
	\nonumber
\,,
\end{eqnarray}
where the ABC is needed to fix the ratio $\rho_0( \kpar ) / \gamma_0( \kpar )$.
For a CC system ($\gamma_0=0$) we recover the well known \cite{garcia,pitarke,ritchie} SP dispersion
$\omega^2=\half\omega_p^2+\omega_p^2 \kpar/2v_b$ with a dominant
linear dispersion at small $\kpar$, i.e.
\begin{equation}
	\text{CC:} \qquad \omega( \kpar ) =
	\frac{\omega_p}{\sqrt{2}} + \frac{ c_b \kpar }{ 2 } + {\cal O}( \kpar^2 )
	\label{eq:CC-approximation}
\end{equation}
%
For the CL system ($\rho_0 = 0$),
we obtain to lowest order as $d \to 0$
a purely quadratic dispersion
$\omega^2 = \half\omega_p^2 + c_s^2 \kpar^2$, featuring the surface speed
of sound $c_s$.
Keeping $d$ finite, and expanding for small $\kpar$, we get
\begin{equation}
	\text{CL:}\qquad \omega( \kpar ) =
	\frac{ \omega_p }{ \sqrt{2} }
	- \frac{c_s \kpar}{2} \tan( {\bar d}/2 ) + {\cal O}( \kpar^2 )
	\,,
	\label{eq:CL-approximation}
\end{equation}
where ${\bar d} = \omega_p d/(\sqrt{2}c_s)$. Note the negative
%
slope, consistent with the charge being outside the bulk
ions (at $z > 0$) \cite{feibelman}.
The divergence at $\tan{\bar d}/2 = \infty$
corresponds to a resonant multimode response \cite{tsuei,sprunger,chiarello,bennett},
which may appear at higher frequencies if the thickness $d$ is large
enough.
Indeed, for $\omega_s = 0$ as assumed here,
we find that this corresponds to a vanishing total layer charge,
$\int_0^d {\rm d}z\, \rho_s( \kpar, z) = 0$: these are surface dipoles or higher
multipoles. They have been observed in alkali metals \cite{tsuei, sprunger,chiarello}
where the charge response outside the bulk is significant. These modes, having
larger linewidths, are not considered in the solutions below.

We note also that in order to get an acoustic plasmon
solution \cite{pitarke,diaconescu} with
$\omega( \kpar ) \to 0$ as $\kpar \to 0$, one would require in Eq.(\ref{e05})
$\gamma_0 d \to -\rho_0 / v_b$ corresponding to opposite signs of the
total bulk ($\rho_0/v_b$) and total surface ($\gamma_0 d$) charges.
A specific ABC is then needed to achieve acoustic plasmons
\cite{pitarke}.
Our model allows for surface scattering between the
two types of charges, a distinct situation.


\paragraph*{{\bf Additional boundary condition via Boltzmann equation.}}
We proceed to derive the ABC by using a Boltzmann transport equation for the charge response \cite{ashcroft,ziman}. This approach is known to be
notoriously difficult when matching conditions at an interface are
involved \cite{ziman}. We have adopted an alternative approach and
consider a bulk half-space terminated by a two-dimensional surface sheet that
represents the charge integrated over the layer $0 < z < d$. The
matching conditions are then replaced by surface scattering that mixes
bulk and surface charges.
At equilibrium, the bulk electron distribution per phase space ${\rm d}^2x {\rm d}z {\rm d}^3p/(2\pi\hbar)^3$ is $f^0({\bf x},z,{\bf p})$ while the surface distribution per phase space ${\rm d}^2x {\rm d}^2q/(2\pi\hbar)^2$
is $f_s^0({\bf x},\qq)$.
The surface charge
sheet involves only the lateral coordinates ${\bf x} = (x, y )$ and a
two-dimensional momentum ${\bf q}$.
We note that this phase space description is possible
even if the surface states are localized, by working with the Wigner
transform \cite{Kirkpatrick,Henkel00}.
In response to a weak electric field, the distributions become $f^0+f, f_s^0+f_s$, with $f,f_s\ll 1$. The transition probability due to bulk impurity scattering per ${\rm d}^3p{\rm d}t$ from state ${\bf p}$ to ${\bf p}'$ is $W({\bf p}',{\bf p})$. Within the Born approximation, or more generally in the presence of time reversal and space inversion, we have $W({\bf p},{\bf p}')=W({\bf p}',{\bf p})$. To 1st order in $f,\EE$, with a bulk velocity ${\bf v}({\bf p})$
\begin{eqnarray}
&&
[\partial_t+\vv({\bf p})\cdot{\bm\nabla}]f({\bf x},z,{\bf p})+e\EE\cdot{\bm\nabla}_{{\bf p}}f^0({\bf x},z,{\bf p})
	\nonumber\\
&&
=
\int_{{\bf p}'}W({\bf p}',{\bf p})[f({\bf x},z,{\bf p}')-f({\bf x},z,{\bf p})]
	\label{e06}
\end{eqnarray}
with the shorthand $\int_{{\bf p}'} = \int\!{\rm d}^3p' / (2\pi\hbar)^3$.
At the surface we have a scattering cross section $W'(\qq',\qq)$ between surface states, as well as scattering of bulk states with $p_z>0$ into the surface
($W_s(\qq,{\bf p})$) and scattering of surface states back into bulk states with $p_z<0$ ($W_s({\bf p},\qq)$). As in the bulk, we assume $W'(\qq',\qq)=W'(\qq,\qq')$. For $W_s(\qq,{\bf p})$ we can assume reflection symmetry only in $x,y$ as well as time reversal, hence $W_s(\qq,{\bf p})=W_s(-\qq,-{\hat {\bf p}})=W_s({\hat {\bf p}},\qq)$, where ${\hat{\bf p}}\equiv (p_x,p_y,-p_z)$. Boltzmann's equation, to first order in $\EE,\,f,\,f_s$ becomes, with a surface velocity ${\bf u}({\bf q})$
\begin{eqnarray}
&&[\partial_t+\uu(\qq)\cdot{\bm\nabla}]f_s({\bf x},\qq)+e\EE\cdot{\bm\nabla}_{\qq}f^0_s({\bf x}, \qq )
	\nonumber\\
&&
=
\int_{\qq'}W'(\qq',\qq)[f_s({\bf x},\qq')-f_s({\bf x},\qq)]\nonumber\\
&& {} +\int_{{\bf p},p_z>0} \kern-2ex
W_s(\qq,{\bf p})
\{
f({\bf x},0,{\bf p}) - f_s({\bf x},\qq)
\nonumber\\
&& {} \qquad
+ f^0_s({\bf x},\qq)[ f({\bf x},0,{\hat{\bf p}}) - f({\bf x},0,{\bf p}) ]
\}
	\label{e07}
\end{eqnarray}
Note that unlike the bulk equation~(\ref{e06}), $f_s^0$ does not cancel.
The final ingredient is a matching condition: the total flux with velocity $\pm v_z$ is matched with the probability of scattering in or out of surface states. Taking
$v_z>0$ and considering the $W_s(\qq,{\bf p})$ symmetries,
\begin{eqnarray}
&&v_z[f({\bf x},0,{\bf p})-f({\bf x},0,{\hat{\bf p}})]
=
\nonumber\\
&&
\int_{\qq}W_s(\qq,{\bf p})\{f({\bf x},0,{\bf p})-f_s({\bf x},\qq)
\nonumber\\
&&
\qquad
{} +f^0_s({\bf x},\qq)[f({\bf x},0,{\hat{\bf p}})-f({\bf x},0,{\bf p})]\}
	\label{e09}
\end{eqnarray}
This relation assumes specular reflection in the absence of surface states, while the latter produce nonspecular reflection in each $v_z$ channel.
Integrating Eq.(\ref{e09}) over ${\rm d}^3p,p_z>0$ yields the total current incoming into the surface, $j_z^{in}({\bf x})$,
and by comparing with Eq.(\ref{e07}) integrated over ${\rm d}^2q$, we get
charge conservation at the surface
\beq{10}
\partial_t{\bar\rho}_s({\bf x})+{\bm\nabla}\cdot{\bf j}_s({\bf x})
= j_z^{in}({\bf x})
\eeq
where ${\bar\rho}_s({\bf x}) = \int_{\qq}f_s({\bf x},\qq)$,
${\bf j_s}( {\bf x} )
= \int_{\qq}\uu(\qq)f_s({\bf x},\qq)$.

To solve the transport problem, we make the $P_1$ approximation,
i.e., on spatial scales large compared to the mean free path, only the
weakest angular dependence in the momentum variable is
maintained \cite{Coppa}. We therefore use the following expansion
\beq{11}
f({\bf x}, z, {\bf p}) &=& \rho({\bf x}, z)f^{(1)}( p )
+ \sum_{ik} j_i( {\bf x}, z ) v_k({\bf p})g_{ik}^{(1)}( p )\nonumber\\
f_s({\bf x},\qq) &=& {\bar\rho}_s({\bf x})f_s^{(1)}( q )
+ {\bf j_s}({\bf x})\cdot\uu(\qq)g_s^{(1)}( q )
\eeq
where $p$, $q$ denote the modulus of the vectors ${\bf p}$, $\qq$.
The reduced symmetry of the surface allows for a diagonal tensor with elements
$g_{xx}^{(1)}( p ) = g_{yy}^{(1)}( p ) \neq g_{zz}^{(1)}( p )$. For elastic scattering, we expect the dependence on $p$, $q$ to be localized near the
corresponding Fermi surface, if the latter is well-defined (the surface states
may be spatially localized).

Integrating the $\vv$ or $\uu$ moment of Eqs.(\ref{e06},\ref{e07}) yields Eq.(\ref{e01}), identifying the plasma frequency as
$\omega_p^2 = 4\pi e^2\int_p\pd{v_i}{p_i} f^0({\bf x},z,{\bf p})$ (assuming diagonal response) which becomes $\omega_p^2 = 4\pi n e^2/m^*$ for a parabolic dispersion with an effective mass $m^*$. Similarly $\omega_s^2
= 4\pi n_s e^2/m^*d$ where $n, n_s$ are equilibrium bulk and surface densities, respectively.
The $\vv$ or $\uu$ moments of Eqs.(\ref{e06},\ref{e07}) yield the
sound velocity
$c_b^2=\int_{{\bf p}}v_i^2f^{(1)}( p )$ and similarly for
$c^2_s$, 
and also identify the bulk and surface relaxation times
\begin{eqnarray}
\frac{1}{\tau} &=& \int_{{\bf p},{\bf p}'}W({\bf p}',{\bf p})v_i^2({\bf p})
g_{ii}^{(1)}( p )
	\nonumber\\
\frac{1}{\tau_s} &=& \int_{\qq,\qq'}W'(\qq',\qq)u_i^2(\qq)g_s^{(1)}( q )
	\nonumber\\
&& \quad {} + \int_{{\bf p},p_z>0}\int_{\qq}W_s(\qq,{\bf p})u_i^2(\qq)
g_s^{(1)}( q )
	\label{e12}
\end{eqnarray}
where $1/\tau$ may in principle be anisotropic.
We also note that the $v_x$ moment of Eq.(\ref{e09}) requires in the
expansion~(\ref{e11}) of the momentum distribution a quadrupole term
$f({\bf x}, z, {\bf p}) \sim v_xv_z$. This term contributes to shear
viscosity at the surface. It is of higher order in velocities, however,
and does not affect the $v_z$ moments that we discuss now.

Taking the $v_z$ moment of the matching
condition~(\ref{e09}), we encounter the following integrals:
\beq{13}
&& \int_{{\bf p}}|v_z|^3g_{zz}^{(1)}( p )
\equiv \alpha_1
	\nonumber\\
&&
\int_{{\bf p},p_z>0}\int_{\qq}W_s(\qq,{\bf p})
v_z [\rho({\bf x},0)f^{(1)}( p )+v_zj_z({\bf x},0)g_{zz}^{(1)}( p )]
	\nonumber\\
&&
\quad \equiv \alpha_2\rho({\bf x},0)+\alpha_3j_z({\bf x},0)
	\nonumber\\
&& \int_{{\bf p},p_z>0}\int_{\qq}W_s(\qq,{\bf p})v_z
f_s^{(1)}( q )
\equiv \alpha_4
\nonumber\\
&&
\int_{{\bf p},p_z>0}\int_{\qq}W_s(\qq,{\bf p})
2
v_z^2 g_{zz}^{(1)}( p ) \,f_s^0({\bf x},\qq)
\equiv\alpha_5 
\eeq
Note that the difference $\alpha_3 - \alpha_5$ involves an integral over
$1 - 2 f_s^0({\bf x},\qq)$, which vanishes if the surface states have an
electron-hole symmetry.

\begin{figure*}
\begin{center}
\includegraphics[width=0.45\textwidth]{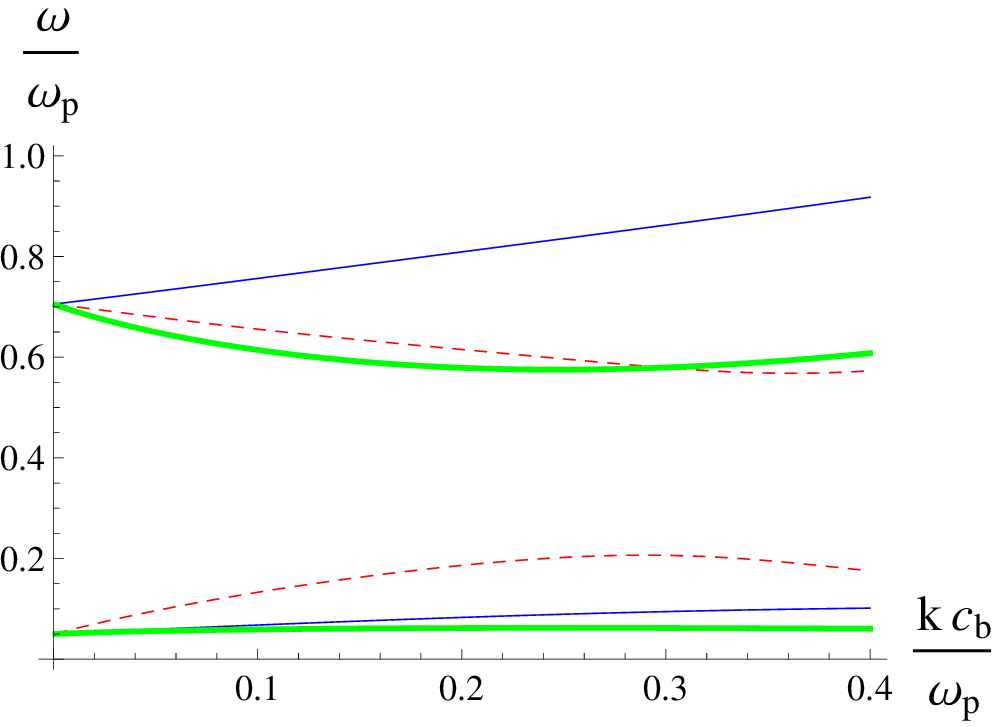}
\hspace*{02mm}
\includegraphics[width=0.45\textwidth]{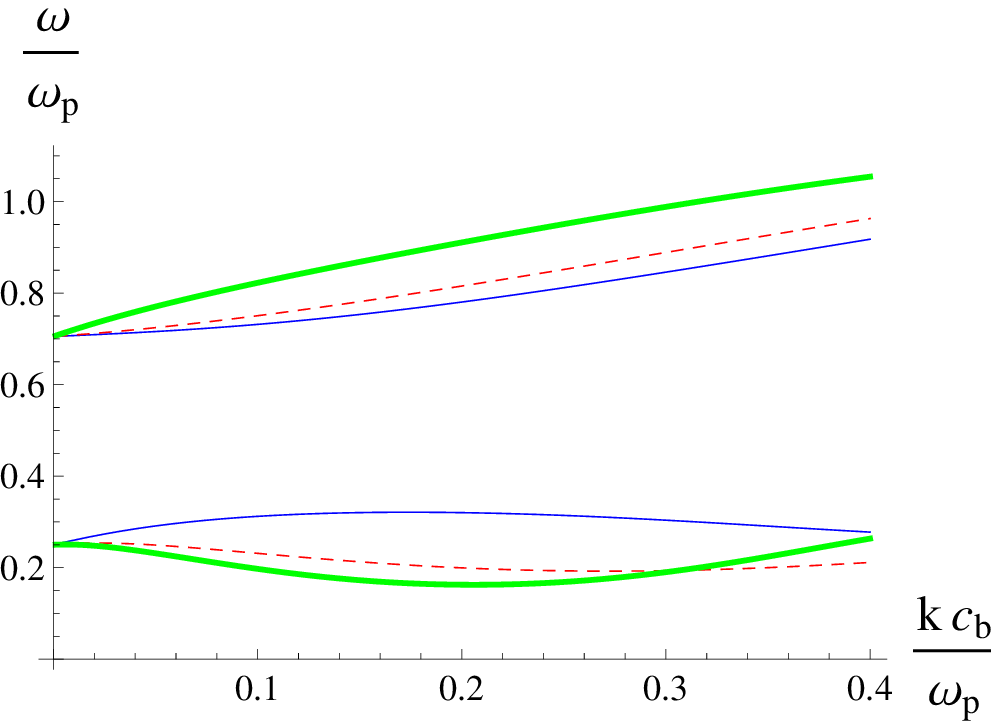}
\end{center}
\caption{(\emph{left})
Surface plasmon dispersion relation $\omega( \kpar )$ for different
values of the surface scattering parameter $\alpha$ in Eq.(\ref{e14}):
$\alpha=0.1$ (thin blue lines), $\alpha=1$ (dashed red lines)
and $\alpha=10$
(thick green lines).
Upper lines starting at $\omega_p / \sqrt{ 2 }$: real part; lower lines:
imaginary part $\times(-1)$.
Other parameters:
$\omega_s=0$, $c_s/c_b=1$, $\omega_p \tau =10$,
$\omega_p \tau_s =10$, $\tau' = \infty$ and ${\tilde d}
= \omega_p d / c_b = 3.5$.
\newline
(\emph{right}) Effects of a nonzero restoring force in the surface layer
(plasma frequency $\omega_s$): $\omega_s=0$ (thin blue curves),
$\omega_s / \omega_p =0.5$ (dashed red curves) and
$\omega_s / \omega_p =0.9$ (thick green curves). The upper curves are the real part and the lower ones are
the imaginary part $\times(-5)$.
The other parameters are $c_s/c_b=\sqrt{3}$, $\omega_p \tau =10$,
$\omega_p \tau_s =10$, $\tau'\omega_p = 1$, $\alpha=0.1$ and ${\tilde d} =7.$
}
\label{fig:SPP-if-omegaS-is-zero}
\label{fig:SPP-if-omegaS-is-nonzero}
\end{figure*}


Putting everything together, the $v_z$ moment of Eq.(\ref{e09}) becomes
the ABC
\beq{14}
j_z^{in}( {\bf x } ) &=&
\alpha c_b \rho_0( {\bf x} ) - \frac{ {\bar\rho}_s( {\bf x} ) }{ \tau' }
\\
\alpha c_b &=& \frac{ \alpha_2 }{ \alpha_1 + \alpha_3 - \alpha_5 }
	\label{eq:def-alpha}
\\
\frac{ 1 }{ \tau' } &=& \frac{ \alpha_4 }{ \alpha_1 + \alpha_3 - \alpha_5 }
\eeq
The two coefficients on the rhs can be interpreted as a probability of
a bulk carrier getting trapped in a surface state (coefficient $\alpha$) and
a desorption rate back into the bulk (rate $1/\tau'$).

%
\paragraph*{{\bf Solutions for the dispersion relation.}}
To make contact with the electromagnetic calculation, we identify
$\bar \rho_s$ with the charge density averaged over the layer
$0 < z < d$, i.e., integrating Eq.(\ref{e01}), we get
${\bar\rho}_s( \kpar ) = [\gamma_0( \kpar ) \sinh v_s d
- \gamma_1( \kpar ) (1 - \cosh v_s d)] / v_s$.
When the boundary condition for the electric field $E_z$
and Eq.(\ref{e04}) for the current $j_z^{in}$ flowing into the layer,
are combined with the ABC~(\ref{e14}), one gets a linear system for
the charge amplitudes $\rho_0( \kpar ), \gamma_0( \kpar ),
\gamma_1( \kpar )$. For the interesting case $\omega_s=0$, this becomes
\beq{16}
\alpha c_b \rho_0( \kpar ) =
[i\frac{c_s^2 \kpar^2-\omega^2}{\omega}+\frac{1}{\tau'}]
\frac{\tanh v_sd}{v_s}\gamma_0( \kpar )
\eeq
and with Eq.(\ref{e05}) yields the SP dispersion
\begin{eqnarray}
&&
c_b\frac{v_b(2\omega^2-\omega_p^2)-\omega_p^2 \kpar}{\omega^2-\omega_p^2}=
\frac{i\alpha}{\omega}\frac{c_s^2\omega_p^2}{\omega^2-c_s^2 \kpar^2+i\omega/\tau'}
	\nonumber\\
&& \:{}\times
\left[v_s \kpar (\frac{\eexp{- \kpar d}}{\sinh v_sd}-\frac{1}{\tanh v_sd})-v_s^2(2\frac{\omega^2}{\omega_p^2}-1)\right]
	\label{e17}
\end{eqnarray}
The parameter $\alpha$ thus measures the relative importance of the CC and CL
amplitudes.
For $\alpha\ll 1$, the dispersion corresponds to a CC oscillation,
Eq.(\ref{eq:CC-approximation}), while for
$\alpha\gg 1$ the dispersion is dominated by the CL,
Eq.(\ref{eq:CL-approximation}).
We recall that a large value of $\alpha$ is
facilitated when electron-hole symmetry holds [$\alpha_3 =\alpha_5$
in Eq.(\ref{eq:def-alpha})].

We display in
Fig.\ref{fig:SPP-if-omegaS-is-zero} numerical solutions for the SP dispersion. On the left the effect of the CC/CL mixing parameter $\alpha$ is shown: the linewidth is relatively small for $\alpha=0.1$ (dominant CC) or $\alpha=10$ (dominant CL), while it is much larger for $\alpha=1$ corresponding to strong CC/CL mixing. This figure is for ${\tilde d}=\omega_pd/c_b=3.5$ demonstrating also a negative initial slope when the CL is significant, i.e. $\alpha=1,10$. When $d$ is smaller, maintaining $\alpha>1$, the linear term is suppressed (Eq. \ref{eq:Feibelman-slope}) and the dispersion becomes purely parabolic, as observed in some cases \cite{sutto,savio1,politano2}. Fig.\ref{fig:SPP-if-omegaS-is-zero}(right) demonstrates the effect of $\omega_s$ and the possibility of generating a local minimum in $\Gamma(k)$, as observed in some cases \cite{chiarello, yu}. This situation is facilitated by $\omega_s\approx \omega_p$ and $\tau'\omega_p=1,{\tilde d}=7$. The effect of $\omega_s$ on the real part is to increase
the linear term in $k$, as discussed in Eq.(\ref{eq:shifted-slope}).

\paragraph*{{\bf Discussion.}}

We developed here a general scheme for solving the SP dispersion at low $k$, allowing for two charge types, CC and CL, both with nonlocal diffusive response. We have solved the notorious ABC problem by a Boltzmann equation that allows for CC/CL surface scattering. The resulting dispersion is consistent with a large variety of data accumulated on pure metals (where the CL corresponds to the spill-out charge), on sputtered systems, disordered surfaces, quantum well surface states, all leading to distinct electron states in a surface layer.

We discuss now the role of the most significant parameters: surface mixing
$\alpha$, layer thickness $d$, surface restoring force $\omega_s^2$, and
how they correspond to observed data.
Consider first the surface mixing: when it is weak (small or large $\alpha$),
it leads to a
relatively small linewidth as in the first eleven lines of Table~\ref{t:table-with-SPdata}. Weak surface mixing corresponds also to dominant CC or CL which is realized e.g. when the dispersion is dominantly linear or quadratic, as indeed in those eleven lines. Cases where the linear and quadratic terms are comparable may (last two lines of Table~\ref{t:table-with-SPdata}) or may not lead to a large linewidth. In the latter case a detailed fit is necessary, and is also demonstrated in Fig. \ref{fig:SPP-if-omegaS-is-zero}(left).

The layer thickness $d$ allows for a negative linear dispersion, as seen in
many metals (Table~\ref{t:table-with-SPdata}).
When $d \to 0$ and CL dominates ($\alpha \gg 1$),
the dispersion is purely quadratic, as indeed
seen in several cases \cite{sutto,savio1,politano2}.

The surface restoring force $\omega_s$ is expected to be small, $\omega_s<\omega_p$, in the case of alkali metals where the spill-out charge is further away from the equilibrium charges. In other cases, a large $\omega_s$ shows the peculiar feature of a minimum in $\Gamma(k)$, as demonstrated in
Fig.\ref{fig:SPP-if-omegaS-is-zero}(right). This minimum, as Al(111) \cite{chiarello} and Ag on Si(111) \cite{yu}, has been discussed in terms of specific band structures. Our approach is an alternative description in terms of the charge distribution. With a few parameters our model is thus able to describe SP dispersion and linewidth on a large variety of metals and surfaces,
as seen over the last decades.





\paragraph*{Acknowledgments.}
We are thankful for stimulating discussions with
E. V. Chulkov, D. Davidov, O. Entin-Wohlman, M. Golosovsky,
F. Guinea, and V. M. Silkin. We also thank E. Eizner
for significant help with data analysis.
This research was supported by a Grant from the G.I.F., the German-Israeli
Foundation for Scientific Research and Development.

\end{document}